\documentclass[9pt,twocolumn,twoside]{osajnl}

\journal{ol} 

\setboolean{shortarticle}{true} 

\usepackage[]{hyperref} 
\hypersetup{pdfauthor={Walasik, Litchinitser},pdftitle={Mode selection using supersymmetry-guided optimization}}

\usepackage{wasysym}

\usepackage{cleveref}

\crefformat{equation}{Eq.~(#2#1#3)}
\crefformat{figure}{Fig.~#2#1#3}

\Crefformat{equation}{Equation~(#2#1#3)}
\Crefformat{figure}{Figure~#2#1#3}

\title{Supersymmetry-guided method for mode selection and optimization in coupled systems}

\author[1]{Wiktor Walasik}
\author[2]{Bikashkali Midya}
\author[2]{Liang Feng}
\author[1,*]{Natalia M. Litchinitser}

\affil[1]{Department of Electrical Engineering, University at Buffalo, The State University of New York, Buffalo, New York 14260, USA}
\affil[2]{Department of Materials Science and Engineering, University of Pennsylvania, Philadelphia,
PA 19104, USA}

\affil[*]{Corresponding author: natashal@buffalo.edu}

\ociscodes{
	(080.6755) Systems with special symmetry; 
	(230.4555) Coupled resonators; 
	(060.2430) Fibers, single-mode; 
	(140.3570) Lasers, single-mode; 
	(140.3325) Laser coupling.
}

\begin{abstract}
Single-mode operation of coupled systems such as optical-fiber bundles, lattices of photonic waveguides, or laser arrays requires an efficient method to suppress unwanted super-modes. 
Here, we propose a systematic supersymmetry-based approach to selectively eliminate modes of such systems by decreasing their lifetime relative to the lifetime of the mode of interest. 
The proposed method allows to explore the opto-geometric parameters of the coupled system and to maximize the relative lifetime of a selected mode.
We report a ten-fold increase in the relative lifetime of the fundamental modes of large one-dimensional coupled arrays in comparison to simple 'head-to-tail' coupling geometries. 
The ability to select multiple supported modes in one- and two-dimensional arrays is also demonstrated. 
\end{abstract}

\setboolean{displaycopyright}{true}

\begin{document}

\maketitle

The concept of supersymmetry (SUSY) originates from the quantum field theory and allows for the treatment of bosons and fermions on equal footing~\cite{Ramond71,NEVEU71,WITTEN81,WITTEN82,Weinberg05}. 
Later, SUSY found applications in quantum mechanics and became a powerful analytical tool for studies of scattering potentials and soliton dynamics. 
SUSY enabled, for instance, analytical treatment of new families of reflectionless~\cite{MAYDANYUK05} and periodic potentials~\cite{DUNNE98,Dunne98a,Khare04}.
A broad review of the mathematical formulation and applications of the SUSY quantum mechanics is presented in~\cite{Cooper01}.

Recently, SUSY was extended to optics~\cite{CHUMAKOV94,Laba14} and helped to address problems of mode control in planar waveguides~\cite{Principe15,Yu17} and optical fibers~\cite{Macho18}, soliton stabilization in parity-time symmetric systems~\cite{Driben11}, Bragg grating design~\cite{Longhi15}, or generation of complex potentials with entirely real spectra~\cite{MILANOVIC02,Miri13a,Rosas-Ortiz15}. 
Various new concepts in SUSY optics were introduced~\cite{Miri13}, including iso-spectral transformations of discrete and continuous potentials~\cite{Miri:14} leading to observation of scattering on discrete SUSY (DSUSY)~\cite{Heinrich:14} potentials and efficient mode-division multiplexing~\cite{Heinrich:14a} in evanescently-coupled photonic-waveguide lattices.

\begin{figure}[!t]
	\centering
	\fbox{\includegraphics[width=0.97\linewidth, clip = true, trim = {0 0 0 0}]{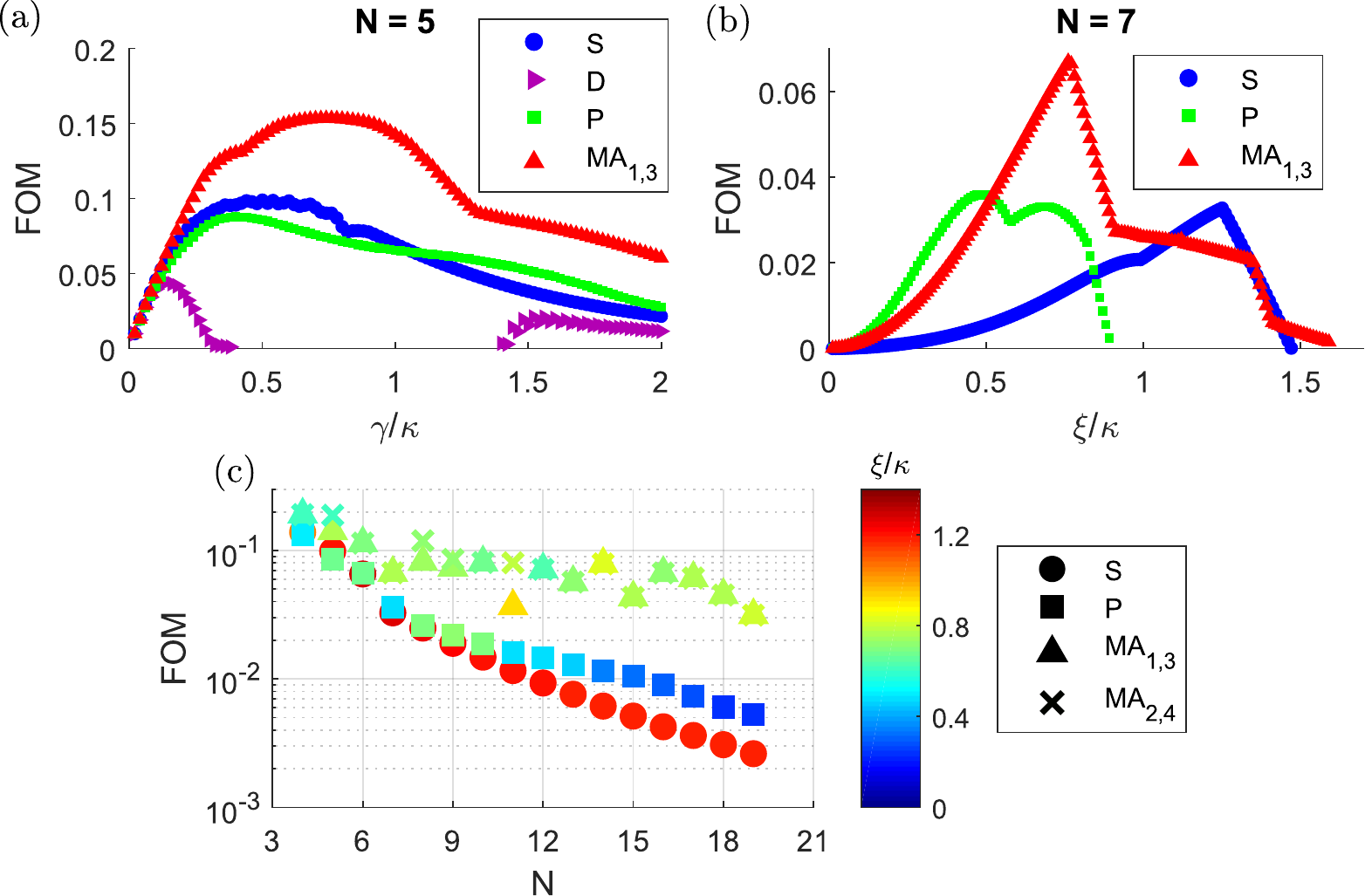}}
	\caption{
		(a)~FOM of the fundamental mode for a 1D array built of 5 identical sites as a function of the loss of the partner array~$\gamma$ for different coupling geometries: 
		single (blue~$\CIRCLE$---S), 
		double (violet~$\blacktriangleright$---D), 
		parallel (green~$\blacksquare$---P), 
		and multi-array (red~$\blacktriangle$---MA$_{N_s,N_m}$). 
		(b)~FOM of the fundamental mode for a 1D array built of 7 identical sites for fixed loss $\gamma= 0.5\kappa$ as a function of coupling constant~$\xi$ between the main and partner arrays for S, P, and MA$_{1,3}$ coupling configurations. 
		(c)~FOM of the fundamental mode as a function of a 1D-array size~$N$.
		The colors of the symbols show the coupling~$\xi$ for which the maximum FOM is achieved.}
	\label{fig:scans}
\end{figure}

Moreover, SUSY was applied to laser design, leading to enhanced transition probabilities in semiconductor quantum-well cascade lasers~\cite{Milanovic96,Tomic97,Bai:08}. 
A DSUSY-based solution of the problem of transverse super-mode selection in integrated lasers~\cite{Kapon84,Wu:07,Ge:15} allowed to suppress undesired modes of one- and two-dimensional (1D and 2D) laser arrays~\cite{El-Ganainy15,Teimourpour16}. 
There, a lossy quasi-iso-spectral SUSY-partner array was coupled with the main array resulting in selective reduction of the lifetime of the unwanted modes (a process known as Q-spoiling). 
In the case of 1D arrays~\cite{El-Ganainy15}, the lossy superpartner was coupled to a terminal resonator of the main array. 
The problem of such a coupling configuration is that its effectiveness decreases rapidly with the increase of the size of the array, as shown in \cref{fig:scans}(c). 
The coupling schemes that we propose here allow one to overcome this issue.
For 2D arrays~\cite{Teimourpour16}, efficient Q-spoiling required coupling of additional resonators, whose parameters were not determined by the DSUSY procedure, and the efficiency was optimized using a trial-and-error method. 
On the contrary, here, we propose an algorithm entirely based on the DSUSY approach that generates coupling configurations that increase the unwanted mode suppression in comparison with the previous approaches.

In this Letter, we develop a systematic method of super-mode selection and optimization in coupled systems.
To increase the attenuation rate of all the unwanted modes of the main array~$A_1$, we couple a lossy array~$A_2$ supporting these modes to the array~$A_1$. 
The modes supported by both~$A_1$ and~$A_2$ will then redistribute in both arrays and become lossy, while the selected mode of interest supported only by~$A_1$ will remain only in the main array and will maintain a long lifetime. 
The problem at hand is how to choose the coupling configuration between the two arrays, to achieve an optimum suppression of the undesired modes.
We methodically explore the space of opto-geometric coupling parameters, such as the loss of the partner array, the coupling configuration, and the coupling strength between the main and partner arrays, in order to find the configuration that maximizes the lifetime of the selected mode relative to the lifetimes of all the other modes. 
We propose a DSUSY-based method relying on maximizing the mode overlap between the main and partner arrays. 
We show that, as a result of the optimization, the range of the single-mode operation for large 1D arrays increases by more than one order of magnitude, as compared to the single-site coupling. 
Finally, we use our method to efficiently select multiple super-modes in 1D and 2D arrays, that allows one to control the pattern and the beating period of their interference. 
These results may find application in control of the transverse-mode profiles of optical-fibers bundles, photonic-waveguide lattices, or integrated laser arrays.    

\begin{figure}[!t]
	\centering
	\fbox{\includegraphics[width=0.97\linewidth, clip = true, trim = {0 0 0 30}]{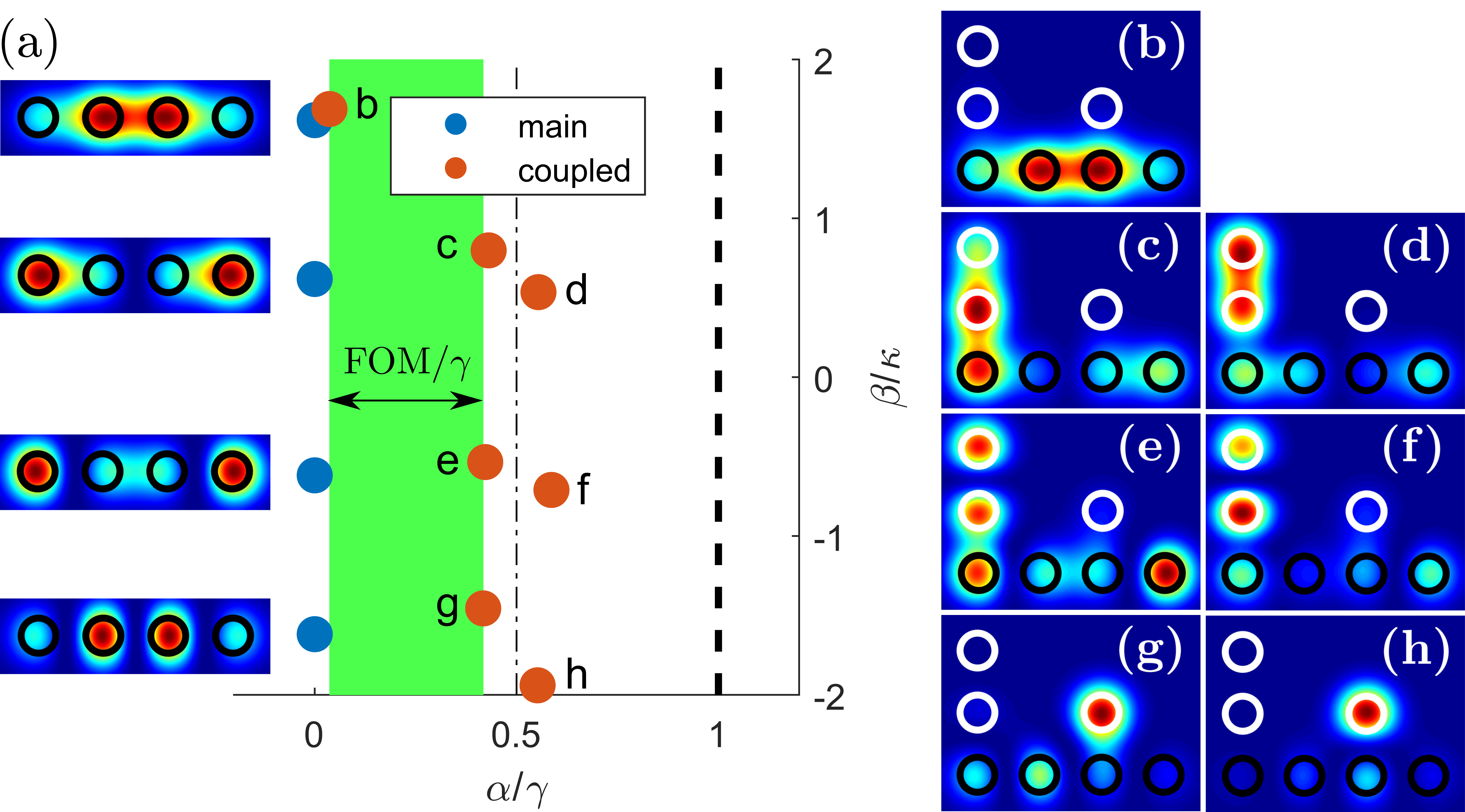}}
	\caption{
		(a)~Complex spectra of a 1D array composed of 4 identical sites (blue) and of the system coupled with a 'multi-array' partner (red). 
		Figure of merit is shaded in green. 
		Insets show intensity distributions of the modes of the main array. 
		(b)--(h)~The intensity distributions of the modes of the coupled systems labeled b--h in (a). 
		Black and white rings mark the location of the sites of the main and  partner arrays, respectively.}
	\label{fig:modes}
\end{figure}

Consider an array~$A_1$ composed of~$N$ identical single-mode coupled fibers, waveguides, or resonators (referred in the following as sites). 
Such a system supports~$N$ super-modes and each mode has the same lifetime~$\tau$, due to the uniform distribution of loss in the system. 
To prioritize a selected mode, we need to increase the attenuation rate $\alpha \propto \tau^{-1}$ of all the other modes (Q-spoiling). 
Here, the difference between the lowest~$\alpha$ among the unwanted modes and~$\alpha$ of the selected mode will be called the figure of merit (FOM) (see green shading in~\cref{fig:modes}(a))\cite{w0}. 

In order to maximize the FOM of a selected mode, we will use the DSUSY-based approach proposed in~\cite{El-Ganainy15,Teimourpour16}. 
The main array~$A_1$ is described using a tight-binding Hamiltonian 
\begin{equation}
\hat{H}_1 =  \sum_{n=1}^N \omega_0 \hat{a}^{\dagger}_n \hat{a}_n + \sum_{n,m=1}^N \kappa_{n,m} \hat{a}^{\dagger}_n \hat{a}_m,
\label{eqn:ham}
\end{equation}
where~$\omega_0$ is the resonant frequency of each site, $\kappa_{n,m}$ denotes the coupling coefficient between the~$n$th and~$m$th sites, and~$\hat{a}^{\dagger}_n$ and~$\hat{a}_n$ are the creation and annihilation operators for photons in the~$n$th site, respectively. 
In the nearest-neighbor coupling approximation, the coupling coefficient $\kappa_{n,m}=\kappa_{m,n}$ is nonzero only if the sites~$n$th and~$m$th are adjacent to each other. 
The Hamiltonian $\hat{H}_1$ can be represented as an $N\times N$ matrix~$H_1$. 

For 1D arrays, the matrix~$H_1$ is symmetric and tridiagonal, and the SUSY-partner array is obtained directly by the DSUSY transformation based on the orthogonal-triangular QR decomposition~\cite{Burden10,Heinrich:14,Heinrich:14a,El-Ganainy15,Teimourpour16}.
For 2D arrays, the Hamiltionian matrix contains two additional diagonals responsible for coupling in the extra dimension, and one more algebraic step is necessary in order to obtain a matrix suitable for the DSUSY transformation. 
The Householder's transformation~\cite{Burden10,Yu:16,Teimourpour16} yields a symmetric tridiagonal $N\times N$ matrix~$H_h$ isospectral to~$H_1$~\cite{w1}.

To remove the mode with the eigenvalue $\tilde{\beta}_n = \beta_n -i\alpha_n$, we need to first shift all the eigenvalues, so that eigenvalue of the mode of interest $\tilde{\beta}_n=0$~\cite{w2}
\begin{equation}
H_{s} = H_{1/h}-\tilde{\beta}_n I_N,
\label{eqn:shift}
\end{equation}  
where~$I_N$ denotes the~$N\times N$ identity matrix. 
Then,~$H_{s}$ is factorized using the QR decomposition
$
H_{s} = QR.
$
Finally, the DSUSY partner is found to be 
\begin{equation}
H_{2} = [RQ]_{\textrm{TL}} + \tilde{\beta}_n I_M.
\label{eq:dsusy}
\end{equation}
The~$RQ$ matrix is block diagonal and each block contains a tridiagonal matrix. 
The top left block of this matrix, denoted by the subscript~$\textrm{TL}$, has the size $M \times M$, and its spectrum is non-degenerate and contains all the eigenvalues of~$H_{s}$ except for $\tilde{\beta}=0$. 
The addition of the second term in \cref{eq:dsusy} brings the spectrum back to the original values. 
$H_2$ describes a 1D array~$A_2$ that is the smallest superpartner of~$A_1$. $A_2$ is quasi-iso-spectral to~$A_1$ because the frequency~$\tilde{\beta}_n$ is removed.

In the following, we study the influence of the opto-geometric parameters of the coupled arrays~$A_1$ and~$A_2$ on the FOM.   
We assume that~$A_1$ is lossless ($\Im m\{\omega_0\}=0$) and that the eigen-frequency of the sites in~$A_1$ is equal to zero ($\Re e\{\omega_0\}=0$). 
The coupling coefficient between all the sites in~$A_1$ has the amplitude~$\kappa$. 
In the optimization, we will study the influence of the following parameters: 
(i)~the loss~$\gamma$ of the partner array (the complex eigen-frequencies of the~$m$th site of $A_2$ is given by $\omega_m - i\gamma$), 
(ii)~the coupling strength between the sites of the main and partner arrays~$\xi$, and
(iii)~the geometric arrangement of the sites corresponding to different coupling configurations.

\begin{figure}[!t]
	\centering
	\fbox{\includegraphics[width=0.8\linewidth]{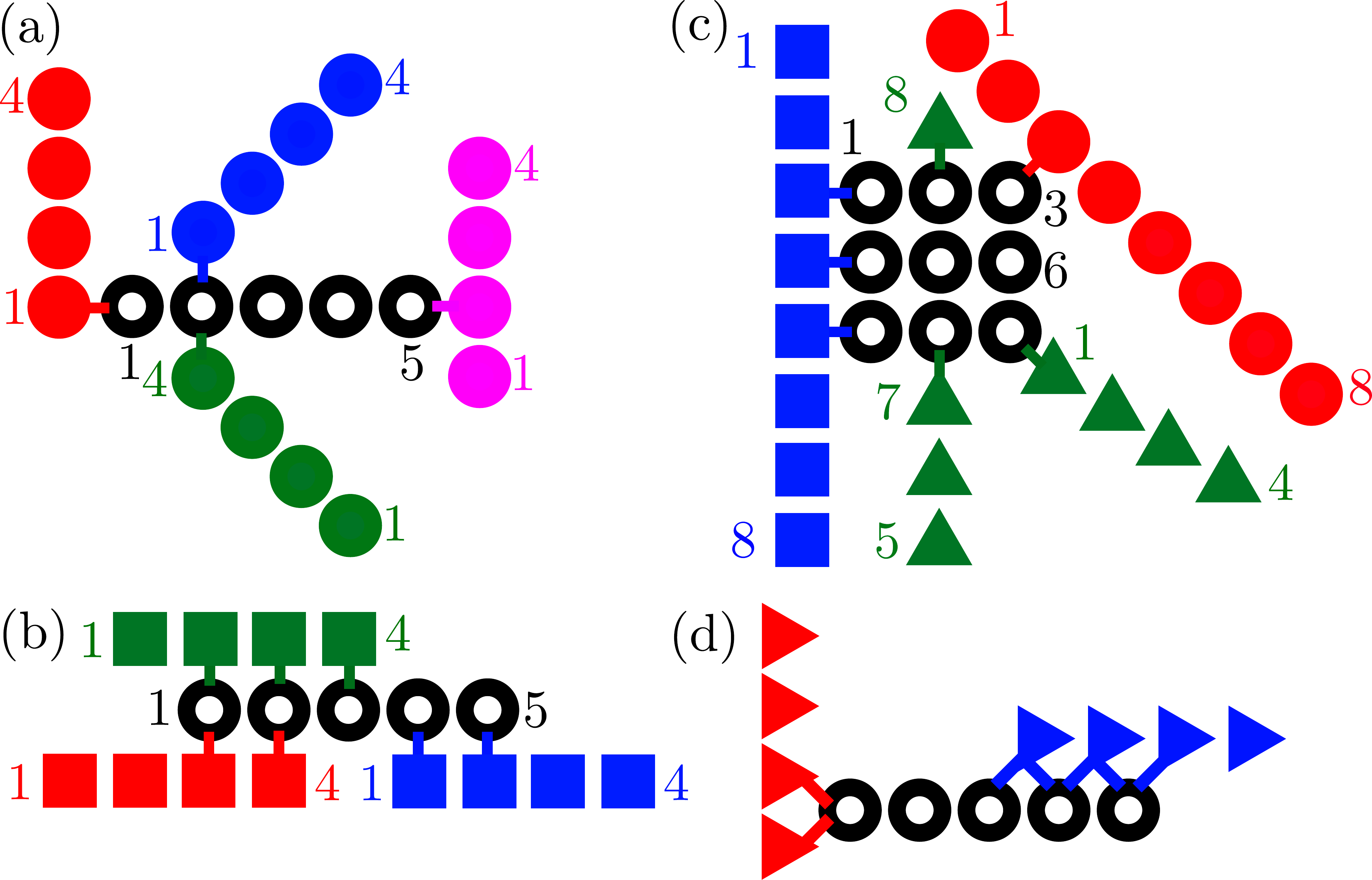}}
	\caption{
		Possible coupling geometries:
		(a)~'single-site' ($\CIRCLE$), 
		(b)~'parallel' ($\blacksquare$), 
		and (d)~'double' ($\blacktriangleright$). 
		(c)~Different coupling configurations for a 2D array including 'multi-array' coupling ($\blacktriangle$) shown in green. 
		Black rings denote sites of the main array while colored symbols show DSUSY-partner arrays. 
		Different colors denote different coupling realizations that are tested one at a time. 	
		Each link between sites form the main and partner arrays represents coupling with magnitude~$\xi$.}
	\label{fig:geom}
\end{figure}

Several realizations of various coupling configurations are schematically illustrated  in \cref{fig:geom}. 
The main array is represented by the black rings and it is coupled to one of the partner arrays represented by symbols of the same color. 
\Cref{fig:geom}(a) shows some of the possible 'single-site' (S) coupling configurations, where one site of~$A_1$ is coupled to one site of~$A_2$. 
Note that due to the symmetry of the main 1D array, coupling with only~$\lceil N/2 \rceil$ sites must be tested. 
On the contrary, in general, the partner array is asymmetric and therefore coupling with each site should be explored. 
\Cref{fig:geom}(b) shows several of the 'parallel' (P) coupling configurations, where~$n$ sites of~$A_1$ are coupled to~$n$ sites of~$A_2$. Here, we assume that the coupling between each pair of sites has the amplitude~$\xi$.
\Cref{fig:geom}(d) shows some of the 'double' (D) coupling configurations in which two neighboring sites from one array are coupled with one site from the second array.
\Cref{fig:geom}(c) shows examples of possible coupling configurations with a 2D array. 
In green, a realization of the 'multi-array' (MA) coupling is illustrated, where the partner array is split and coupled to multiple sites of the main array. 

In contrast to the S, P, and D coupling schemes, that explore all the geometric configurations, the MA scheme tests only the configurations with a significant mode overlap between the main and partner arrays.
Let's consider a 1D array with~$N=4$ sites, illustrated in~\cref{fig:modes}.
First, we generate the partner array~$A_2$ with the frequency of the fundamental super-mode of~$A_1$ removed from its spectrum, using the DSUSY procedure given by Eqs.~(\ref{eqn:shift}) and (\ref{eq:dsusy}).
Then, based on the mode distribution, we choose through which site each of the remaining modes are coupled most efficiently. 
From~\cref{fig:modes}(a), we see that the~$2^{\textmd{nd}}$ and~$3^{\textmd{rd}}$ mode are mostly localized in the terminal sites (1 and 4) and that the~$4^{\textmd{th}}$ mode resides mostly in the middle sites (2 and 3). 
If we denote the number of possible coupling locations for the~$m$th mode as~$L_m$ then the number of all possible sets of location--mode pairs is given by $\prod_{m=2}^N L_m$. 
In~Figs.\ref{fig:modes}(b)--\ref{fig:modes}(h), a set where the~$2^{\textmd{nd}}$ and~$3^{\textmd{rd}}$ modes are coupled via site 1 and the~$4^{\textmd{th}}$ mode is coupled via site 3 is illustrated. 
For each set, we create the arrays attached to the selected sites. 
The arrays are created by cascaded DSUSY transformations of the array~$A_2$. 
At each step, one of the modes not coupled through a given site is removed and the resulting array is then attached to this site. 
In the example shown in~\cref{fig:modes}, one array is created by removing the frequency of the~$4^{\textmd{th}}$ mode from~$A_2$, and the second array (in this case a single site) is created by removing the frequencies of the~$2^{\textmd{nd}}$ and~$3^{\textmd{rd}}$ modes from~$A_2$.
As the generated arrays are in general asymmetric, they can be coupled to the chosen site with either one of their terminal sites.
Using the resulting MA configuration, we obtained the $\textrm{FOM}/\gamma=0.38$ for this system by efficiently coupling the unwanted modes to the lossy partner arrays, and leaving the fundamental mode unaffected and residing in the lossless main array (see Figs.~\ref{fig:modes}(b)--(h)).

For the small array shown in~\cref{fig:modes}, the number of possible MA configuration is small and all of them can be easily tested. 
For large structures supporting many modes with more complex spatial distributions, the number of possible coupling configurations becomes very large. 
Therefore, it is advantageous to use some constraints to limit the number of possible configurations. 
Here, we limit the maximum number of arrays consisting only of a single site (denoted by~$N_s$) and the maximum number of arrays into which~$A_2$ can be split ($N_m$).
Such configurations will be denoted by MA$_{N_s,N_m}$.
If the second constraint is impossible to fulfill, then~$N_m$ is set to the minimum number of arrays for which efficient coupling of all the modes of interest is possible.
Further reduction of the number of possible configurations can be achieved considering the symmetry of the main array.
For example, for rectangular arrays, coupling with the sites located only along two adjacent edges can be tested.

\Cref{fig:scans}(a) shows the maximum FOM (for optimized coupling~$\xi$) of the fundamental mode for a 1D array built of $N=5$ sites obtained for different coupling configurations as a function of the loss of the partner array~$\gamma$. 
We see that the largest FOM is obtained using an MA configuration.
Moreover, the maximum FOM for the S, P, and MA configurations is achieved for~$\gamma \in [0.4\kappa, 0.8\kappa]$.
For $\gamma<0.4\kappa$, the loss is simply not high enough to give a large FOM.
On the contrary, for $\gamma > \kappa$, the large difference in the loss level in~$A_1$ and~$A_2$ leads to decreased coupling efficiency. 
Finally, we observe that for D configurations, it is impossible to achieve a positive FOM for $\gamma \in [0.4\kappa, 1.4\kappa]$.
The reason for that is that for higher-order modes, the field in the neighboring sites is out-of-phase.
Coupling from two out-of-phase sites to a single site, as shown in~\cref{fig:geom}(d), is not efficient.

Smooth sections of the FOM($\gamma$) curves correspond to a fixed configuration yielding the largest FOM.
Abrupt changes of the slope of FOM($\gamma$) reflect the change of the geometric arrangement providing the largest FOM.
The dependencies of FOM on~$\gamma$ for other sizes of 1D arrays show similar behavior to that illustrated in~\cref{fig:scans}(a) for $N=5$.
Therefore, to maximize the FOM, in the following, we use $\gamma = 0.5\kappa$.

\Cref{fig:scans}(b) shows the dependence of the fundamental-mode FOM on the coupling strength between the arrays~$\xi$ for the S, P, and MA configurations for a 1D array composed of $N=7$ sites. 
For $\xi<0.5\kappa$, the P configurations provide the largest FOM. 
For $\xi\in[0.5\kappa, \kappa]$, the MA configurations are optimal, while for larger values of~$\xi$, the S configurations prevail. 
Figures \ref{fig:scans}(a) and \ref{fig:scans}(b) show the necessity for a careful optimization of the loss~$\gamma$ and the coupling~$\xi$ in order to maximize the FOM.

\Cref{fig:scans}(c) shows the dependency of the FOM of the fundamental mode on the size~$N$ of a 1D array.
We observe that the efficiency provided by the S configurations decreases rapidly with the increase of the size of~$A_1$, as the localization of the modes in the center of the array increases.
The P configuration allows to double the FOM for arrays with $N>15$, while the MA configurations, with only two single sites and a total of four split arrays, allow to increase the FOM by an order of magnitude. 
Finally, we observe that for the S configurations, the optimum couping $\xi \approx 1.2\kappa$; for MA scheme, $\xi\in[0.6\kappa, 0.9\kappa]$; and for the P configurations, the optimum~$\xi$ decreases monotonically with the increase of~$N$. 
Moreover, for P configurations, the number of coupled site-pairs increases with the increase of~$N$.
These effects cause the global coupling between~$A_1$ and~$A_2$, computed as~$\xi$ multiplied by the number of coupled site-pairs, to remain almost constant at the level of $\approx 1.2\kappa$, regardless of the coupling configuration and the array size~$N$.

\begin{figure}[!t]
	\centering
	\fbox{\includegraphics[width=0.97\linewidth, clip = true, trim = {0 0 10 0}]{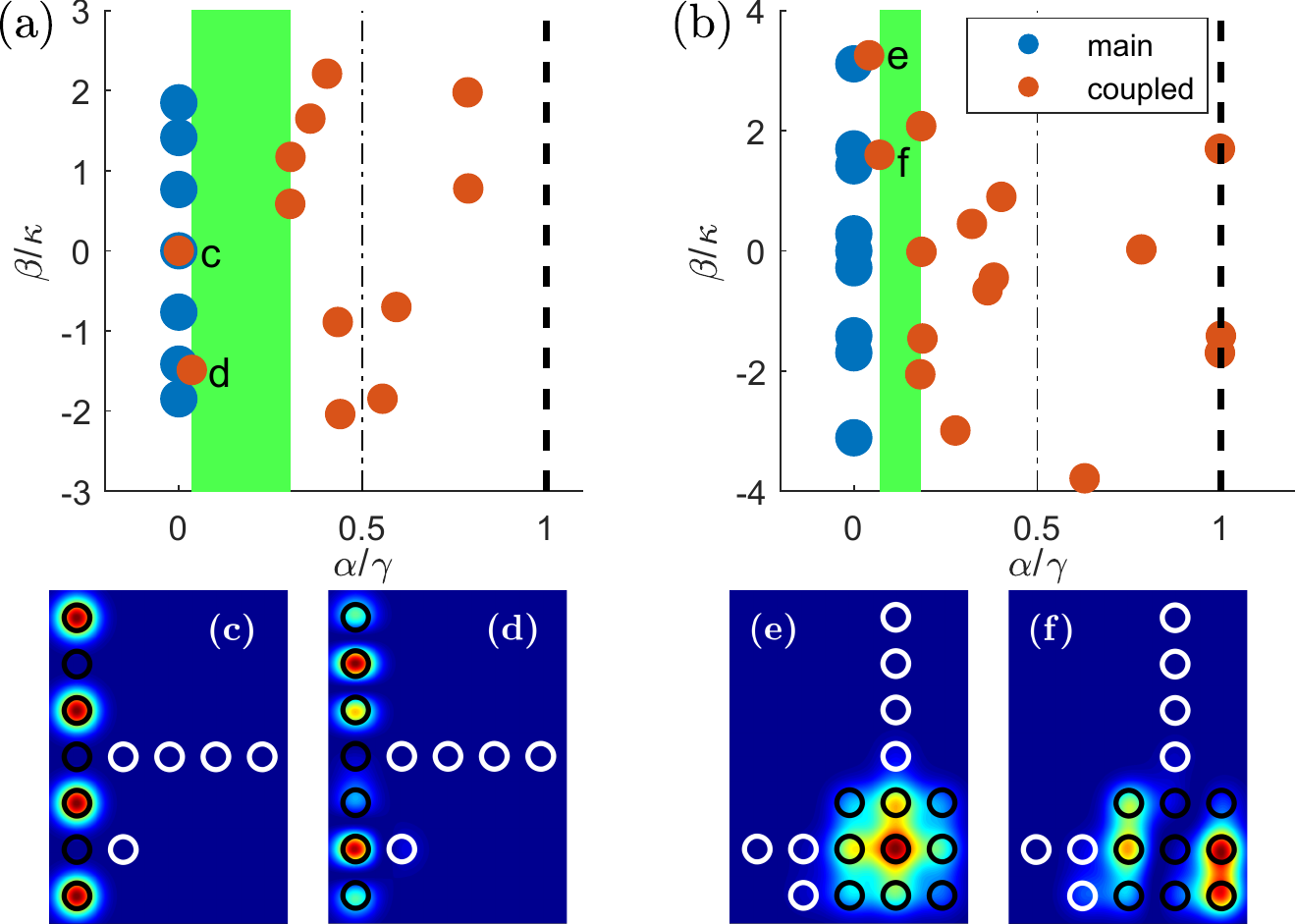}}
	\caption{
		Selection of more than one mode. 
		(a), (c), (d)~The~$4^{\textrm{th}}$ and~$6^{\textrm{th}}$ modes of the 1D array build of $N=7$ sites are chosen. 
		(b), (e), (f)~The~$1^{\textrm{st}}$ and~$3^{\textrm{rd}}$ modes of the 2D $3\times3$ array are selected. 
		Panels (a) and (b) show the spectra of the coupled systems, and (c)--(f) show the intensity distributions for the selected modes.}
	\label{fig:multi}
\end{figure}

Finally, we use our optimization procedure to design arrays supporting multiple arbitrarily chosen super-modes. \Cref{fig:multi} shows two examples of such structures. 
In~\cref{fig:multi}(a), the~$4^{\textrm{th}}$ and~$6^{\textrm{th}}$ modes from a 1D arrays with $N=7$ sites are selected with the $\textrm{FOM}/\gamma=0.3$.
In~\cref{fig:multi}(b), our optimization procedure is used to select the~$1^{\textrm{st}}$ and~$3^{\textrm{rd}}$ modes of the 2D $3\times 3$ array.
Here, the degeneracy of the super-modes is lifted by introducing the asymmetry between the horizontal and vertical coupling constants $\kappa_v = 1.2 \kappa_h = 1.2\kappa$.  
In this case, the largest $\textrm{FOM}/\gamma=0.11$ is offered by one of the MA configurations.
Figures~\ref{fig:multi}(c)--\ref{fig:multi}(f) show that the selected mode profiles are slightly perturbed but remain strongly localized in the main array.
The ability to efficiently select an arbitrary set of super-modes in large arrays might be used to optimize the mode profile in coupled-waveguide systems or to control the far-field distribution or directionality of integrated resonators or waveguides.

In summary, we have developed a systematic approach to optimize the performance of large-scale coupled systems such as optical-fiber bundles, photonic waveguide arrays, or large-area integrated lasers. 
Our supersymmetry-based technique allows to select arbitrarily chosen mode and increase its lifetime with respect to the remaining modes. 
We have swept the space of opto-geometric parameters of the coupled system, such as the loss of the partner array, the coupling configuration, and the coupling strength between the main and partner arrays. 
For selected examples, we have illustrated the importance of a proper choice of the loss and the coupling strength.
Moreover, we have shown that our 'multi-array' coupling scheme allows to increase the figure of merit by at least one order of magnitude, due to the choice of coupling geometry guided by the mode overlap in the main and partner arrays.
Last but not least, we have shown that our method allows for selection of multiple modes both in one- and two-dimensional coupled systems.
Thus, the mode optimization scheme developed here provides a convenient tool for experimental design of coupled systems.

\vspace{1em}

\noindent
\textbf{Funding.} Army Research Office (1142710-1-79506).


\begin{thebibliography}{10}
\bibitem{w0}
	From the experimental viewpoint, the quantity of interest is the ratio between
	the intensities of the selected mode ($s$) and the strongest among the
	unwanted modes ($u$): $Q = I_s/I_u \propto \exp[-2(\alpha_s-\alpha_u)t]$.
	Therefore, maximizing $\textrm{FOM} = \alpha_u-\alpha_s$ maximizes $Q$ at any
	moment of time $t$.
\bibitem{w1}
	Both the Householder's method and the DSUSY transformation described later, may
	result in negative coupling constants~$\kappa_{n,m}$ that are difficult to
	realize in the coupled system. Here, we change the sign of these coupling
	constants to positive. This modification does not affect the spectrum of the
	Hamiltonian.
\bibitem{w2}
	This step has a direct analogy in the SUSY transformation for continuous
	systems. There, a Hamiltonian~$\hat{H}$ is decomposed into the product of two
	operators $\hat{H} = \hat{A}^{\dagger}\hat{A}$, where the operator~$\hat{A}$
	annihilates the ground state~$\psi_0$ whose energy is removed from the SUSY
	spectrum: $\hat{A}\psi_0 = 0$~\cite{Cooper01}. In the DSUSY procedure, the
	matrix operation $R\vec{\phi}_n = \vec{0}$, where~$\vec{\phi}_n$ denotes the
	eigen-vector corresponding to the~$n$th mode.
\end{thebibliography}


\end{document}